\documentclass[11pt]{article}
\usepackage{graphicx}
\usepackage{amsmath}

\newcommand{\BABARPubYear}    {06}

\newcommand{\SLACPubNumber} {12013}

\input babarsym

\long\def\inst#1{\par\nobreak\kern 4pt\nobreak
    {\it #1}\par\vskip 10pt plus 3pt minus 3pt}

\begin{document}
{\pagestyle{empty}

\begin{flushright}
\babar-CONF-\BABARPubYear/009 \\
SLAC-PUB-\SLACPubNumber \\
\end{flushright}

\par\vskip 5cm

\begin{center}
\Large \bf Observation of the decays $\Bm \to D^{(*)+}_s K^- \pi^-$
\end{center}
\bigskip

\begin{center}
\large The \babar\ Collaboration\\
\mbox{ }\\
\today
\end{center}
\bigskip \bigskip

\begin{center}
\large \bf Abstract
\end{center}
We report first observations of the decays $\Bm\to D_s^{(*)+}  K^-   \pi^-$, 
using 292 \invfb\ of data collected at
the $\Upsilon(4S)$ resonance energy by the \babar\ detector at the \pep2\
$e^+ e^-$ collider.
The branching fractions are measured to be
${\cal B}(\Bm \to D^+_s K^- \pi^-) = (1.88 \pm 0.13 \pm 0.41) \cdot 10^{-4}$
and ${\cal B}(\Bm \to D^{*+}_s K^- \pi^-) = (1.84 \pm 0.19 \pm 0.40) \cdot 10^{-4}$.

\vfill
\begin{center}

Submitted to the 33$^{\rm rd}$ International Conference on High-Energy Physics, ICHEP 06,\\
26 July---2 August 2006, Moscow, Russia.

\end{center}

\vspace{1.0cm}
\begin{center}
{\em Stanford Linear Accelerator Center, Stanford University, 
Stanford, CA 94309} \\ \vspace{0.1cm}\hrule\vspace{0.1cm}
Work supported in part by Department of Energy contract DE-AC03-76SF00515.
\end{center}

\newpage
} 

\begin{center}
\small

The \babar\ Collaboration,
\bigskip

%
{B.~Aubert,}
{R.~Barate,}
{M.~Bona,}
{D.~Boutigny,}
{F.~Couderc,}
{Y.~Karyotakis,}
{J.~P.~Lees,}
{V.~Poireau,}
{V.~Tisserand,}
{A.~Zghiche}
\inst{Laboratoire de Physique des Particules, IN2P3/CNRS et Universit\'e de Savoie,
 F-74941 Annecy-Le-Vieux, France }
% ...

\end{center}\newpage

\section{INTRODUCTION}
\label{sec:Introduction}

\begin{figure}[h]
\begin{center}
\includegraphics[width=10.5cm, height=5.5cm, clip=true]{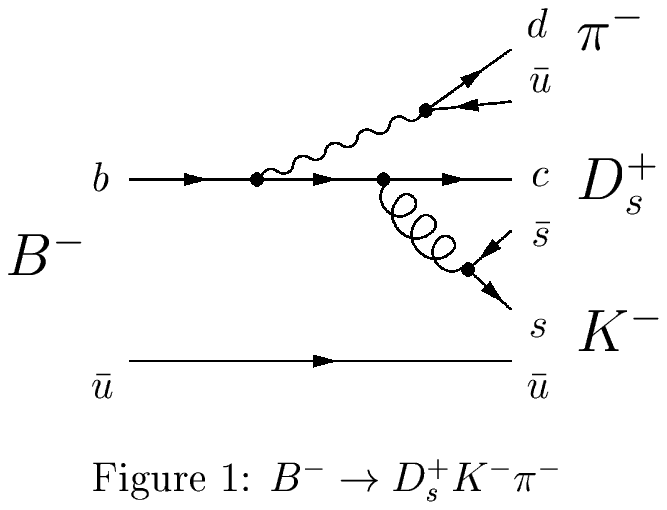}
\end{center}
\caption{Feynman diagram for $\Bm \to D_s^+ K^- \pi^-$. 
}
\label{fig:intro-1}
\end{figure}

First evidence for 
so-called inclusive {\it flavor correlated production} 
of $D_s^+$ in $B^-$ decays was
reported recently~\cite{elba05-couderc} with a branching fraction of 
${\cal B}(\Bm \to D_s^+ X) = (1.2\pm 0.4)\%$~\cite{charge}. 
These decays are mediated by a $b\to c$ 
quark transition and require at least three final state particles,
including the production of an $s\bar{s}$ pair from the vacuum (
$s\bar{s}$ ``popping''). An example for a three-body $\Bm$
decay with a  $D_s^+$ in the final state is $\Bm \to D_s^+ K^-
\pi^-$. The corresponding $\bar{B}^0$ decay is $\bar{B}^0\to D_s^+
\bar{K}^0 \pi^-$. The Feynman diagram for $\Bm \to D_s^{(*)+}  K^- \pi^-$ decays is shown in
Fig.~\ref{fig:intro-1}. In case of $\bar{B}^0\to D_s^+
\bar{K}^0 \pi^-$, an additonal contribution from a $W$-exchange 
diagram with $s{\bar s}$ and $d{\bar d}$ popping may exist. 
If we replace the $\pi^-$ in Fig.~\ref{fig:intro-1}, which comes from the
hadronization of the $W^-$ boson with a $K^-$, we get the
Cabibbo-suppressed decays $B^-\to D_s^+ K^- K^-$ and $\bar{B}^0\to D_s^+
\bar{K}^0 K^-$. It is interesting to note
that the final state $D_s^+ \bar{K}^0 K^-$ can also be reached from a
$B^0$ decay. In this case the decay is mediated by a $b\to u$ quark
transition, but the $W$ hadronization is not Cabibbo-suppressed. Thus a
$\bar{B}^0$ can either decay directly to $D_s^+ \bar{K}^0 K^-$ or via 
$B^0\bar{B}^0$ mixing followed by $B^0\to D_s^+ \bar{K}^0 K^-$. 
The interference between the two decay amplitudes for decay with and
without $B^0\bar{B}^0$ mixing leads to a time-dependent \CP-asymmetry
that is sensitive to $\sin(2\beta +\gamma)$. 
In case the contribution from the higher $D^{**}$ resonances decaying
into $D^+_s {\bar K}$ turns out to be large, it may also be interesting
to measure the resonant parameters independently from the analysis
using $B \to {\bar D} \pi \pi$ decays~\cite{belle-dpipi}.

No exclusive $B^- \to D_s^{(*)+}X$ or $\bar{B}^0 \to D_s^{(*)+}X$ decay
mode has hitherto been observed. Limits on the branching fractions
from the analyses by other experiments are listed in Table~\ref{tab:previous-limits}. 
In this paper we report the first measurement of the decay modes  $\Bm\to D_s^{(*)+}  K^-
\pi^-$.   

\begin{table}
\caption{\label{tab:previous-limits} Upper limits from
  ARGUS~\cite{argus-limit} and CLEO~\cite{cleo-limit} on 
$\Bm \to D_s^{(*)+}  K^- \pi^-$ branching fractions.}

\begin{center}
\begin{tabular}{llc}
Experiment & Decay Mode & Upper limit ($@ 90\%~C.L.$)\\[3pt]
\hline\\[-7pt]
ARGUS~~~~~~~~~~~~~~~~~ & $\Bm \to D_s^+ K^- \pi^-$~~~~~~~~~~ & ~~~~~~~~~~$8\times 10^{-4}$~~~~~~~~~~~\\[4pt]
      & $\Bm \to D_s^{*+} K^- \pi^-$ & $12\times 10^{-4}$\\[4pt]
\hline\\[-7pt]
CLEO  & $\Bm \to D_s^+ K^- \pi^-$ & $5\times 10^{-4}$\\[4pt]
      & $\Bm \to D_s^{*+} K^- \pi^-$ & $6.8\times 10^{-4}$\\[3pt]    
\hline
\end{tabular}
\end{center}
\end{table}

\section{THE \babar\ DETECTOR AND DATASET}
\label{sec:babar}

The analysis uses a sample of approximately $292~$\invfb, which corresponds to 
about 324 million \FourS decays into
$B \Bbar$ pairs collected 
with the \babar\
detector at the PEP-II~\cite{pep2} asymmetric-energy $B$-factory. The \babar\
detector is described elsewhere~\cite{ref:babar} and only the components crucial 
to this analysis are summarized here. Charged particle tracking is
provided by a five-layer silicon vertex tracker (SVT) and a 40-layer
drift chamber (DCH). For charged-particle identification, ionization
energy loss ($dE/dx$) in the DCH and SVT, and Cherenkov radiation
detected in a ring-imaging device are used. 
Photons are identified and
measured using a thallium-doped CsI-crystal electromagnetic calorimeter.
These systems are located inside a 1.5~T 
solenoidal superconducting magnet. We use GEANT4~\cite{geant} software to
simulate interactions of particles traversing the \babar\ detector,
taking into account the varying detector conditions and beam
backgrounds. 

\section{ANALYSIS METHOD}
\label{sec:Analysis}

The optimal selection
criteria as well as the probability density distributions of selection
variables are determined by a blind analysis based on Monte Carlo (MC)
simulation of both signal and background. 
For the calculation
of the expected signal yield we assume ${\cal B}(\Bm \to D_s^{(*)+} K^- \pi^-)$ to be 
$10^{-4}$ ({\it i.e.} about $10\%$ of the measured ${\cal B}(\Bm \to D^{+} \pi^- \pi^-)$~\cite{belle-dpipi}).
We use MC samples of our signal modes and, to simulate background, 
inclusive samples of 
$B^+B^-$ (784~fb$^{-1}$), 
$B^0 \Bbar^0$ (774~fb$^{-1}$),   
$c\bar{c}$ (247~fb$^{-1}$), and   
$q\bar{q},~q = u,d,s$ (246~fb$^{-1}$). In addition, we use large samples of simulated
events of rare background modes 
which have final states similar to the signal.  
We have verified that our MC correctly describes the data by comparing
distributions of various selection variables. 

Candidates for \Ds mesons are reconstructed in the modes $\Ds\to
\phi\pi^+$, $\Kstarzb K^+$, and $\KS K^+$, with $\phi\to K^+K^-$,
$\Kstarzb\to K^-\pi^+$ and $\KS\to\pi^+\pi^-$. 
The \KS candidates
are reconstructed from two oppositely-charged tracks, 
that come from a
common vertex displaced from the $e^+e^-$ interaction point.
We require the significance of this displacement (measured flight distance
divided by an estimated error) to exceed 2.
All other tracks are required to originate less than 1.5~cm away from the
$e^+e^-$ interaction point in the transverse plane and less than 10~cm
along the beam axis.
Charged kaon candidates must satisfy kaon identification criteria
that are typically around 92\% efficient, depending on momentum and
polar angle, and have a pion misidentification rate at the 5$\%$ level.
The $\phi\to K^+K^-$, 
$\Kstarzb\to K^-\pi^+$ and $\KS\to\pi^+\pi^-$ candidates are required to have invariant
masses close to their nominal masses
(we require the absolute differences
between their measured masses and the nominal values~\cite{pdg} to be in the range
$\pm 15$~\mev,
$\pm 50$~\mev and $\pm 10$~\mev, respectively).
The polarizations of the \Kstarzb and $\phi$ mesons in the \Ds decays are
employed to reject backgrounds through the use of the helicity
angle $\theta_H$, defined as the angle between the $K^-$ momentum vector and the
direction of flight of the \Ds in the 
\Kstarzb or $\phi$ rest frame. 
The \Kstarzb and $\phi$ candidates are required to have $|\cos\theta_H|$ greater than 0.5. 

The $D^{*+}_s$ candidates are reconstructed in the mode $D^{*+}_s \ra D^+_s \gamma$.
The photons are accepted if their energy is greater than 100~\mev.
The \Ds and $D^{*+}_s$ candidates are
required to have invariant masses in the interval $[-10,10]$~\mevcc (for \Ds) and 
$[-15,10]$~\mevcc (for $D^{*+}_s$)
from their nominal values~\cite{pdg} (the \Ds mass resolution
is around 6~\mevcc, and the asymmetric mass cut on $D^{*+}_s$ has an efficiency of about $90\%$). 
All $D^+_s$ candidates are mass-constrained.
The invariant mass of the
$D^{*+}_s$ is calculated after a mass constraint on the daughter $D^+_s$ has been applied.
Subsequently, all $D^{*+}_s$ candidates are subjected to a mass-constrained fit. 

We also require that photons from $D^{*+}_s$ are inconsistent 
with $\pi^0$ hypothesis when combined with any other photon having an energy
greater than 150~\mev in the event
(the $\pi^0$ veto window is $\pm 10$~\mevcc). 
Finally, the $\Bm$ meson candidates are formed using the reconstructed
combinations of $D^+_s K^- \pi^-$ and $D^{*+}_s K^- \pi^-$.

The background from continuum $q\bar{q}$ production (where $q = u,d,s,c$) is
suppressed based on the event topology. We calculate the angle ($\theta_T$)
between the thrust axis of the $B$ meson candidate and the thrust axis
of all other particles in the event in the center-of-mass frame (c.m.). In this frame,
$B\Bbar$ pairs are produced approximately at rest and have a 
uniform $\cos\theta_T$ distribution. In contrast, $q\bar{q}$ pairs are
produced in the c.m.\ frame with high momenta, which results in a
$|\cos\theta_T|$ distribution peaking at 1. $|\cos\theta_T|$  is required to be smaller than
0.8. In addition, the
ratio of the second and zeroth order Fox-Wolfram
moments~\cite{fox-wolfram} must be less than 0.3.

We extract the signal using the kinematical variables $\mes =
\sqrt{E_{\rm b}^{*2} - (\sum_i {\mathbf p}^*_i)^2}$ and $\Delta E =
{\sum_i}{\sqrt{m_i^2+{\mathbf p}_i^{*2}}} - E_{\rm b}^*$, where $E_{\rm b}^*$ is
the beam energy in the c.m.\ frame, ${\mathbf p}^*_i$ is the c.m.\
momentum of the daughter particle $i$ of the $\Bm$ meson candidate, and
$m_i$ is the mass hypothesis for particle $i$. For signal events, \mes
peaks at the $\Bm$ meson mass with a resolution of about 2.6~\mevcc and
$\Delta E$ peaks near zero with a resolution of 13~MeV, indicating that
the $\Bm$ candidate has a total energy consistent with the beam energy in
the c.m.\ frame. The $\Bm$ candidates are required to have $|\Delta E|<
25\ \mev$ (around $2\sigma$ of the signal $\Delta E$ resolution) 
and $\mes > 5.2~\gevcc$.

The fraction of events with multiple $\Bm$ candidates is estimated using the
MC simulation and found to be around $3\%$ for $D^+_s K^- \pi^-$
and $9\%$ for $D^{*+}_s K^- \pi^-$ combinations. In each event with
more than one $\Bm$ candidate that passes the selection requirements, we select
the one with the lowest $|\Delta E|$ value.

After all selection criteria are applied, we estimate the $\Bm$
reconstruction efficiencies, excluding the subsequent branching fractions
(see Table~\ref{tab:eff}).

\begin{table}[!h]
\begin{center}
\caption{Reconstruction efficiencies for $\Bm \ra D^{(*)+}_s K^- \pi^-$ decays (excluding the subsequent
branching fractions).}
\label{tab:eff}
\vspace{\baselineskip}
\begin{tabular}{ c c c c c }
 Decay mode  & $D_s^+ \ra \phi \pi^+$~ & $D^+_s \ra \Kstarzb K^+$~ & 
$D^+_s \ra \KS K^+$ \\[4pt]
\hline\\[-7pt]
$\Bm \ra D^+_s K^- \pi^-$ & 11.0$\%$ & 7.0$\%$ & 10.0$\%$ \\
\\[-7pt]
$\Bm \ra D^{*+}_s K^- \pi^-$ & 5.3$\%$ & 3.4$\%$ & 4.8$\%$ \\
\\[-7pt]
\hline
\end{tabular}
\end{center}
\end{table}

Background events that pass these selection criteria are represented by
approximately equal amounts of $q\bar{q}$ continuum and $B{\bar B}$ events. We parametrize
their \mes distribution by a
threshold function~\cite{Argus}: 
$$
f(m_{\rm ES}) \sim m_{\rm ES} \sqrt{1-x^2} {\rm exp}[-\xi (1-x^2)],
\label{eq:argus}
$$
where $x = 2 m_{\rm ES}/\sqrt{s}$, $\sqrt{s}$ is the total energy
of the beams in their center of mass frame, and $\xi$ is the
fit parameter. 

\begin{figure}[ht]
\begin{center}
\includegraphics[width=8.1cm, height=5.5cm]{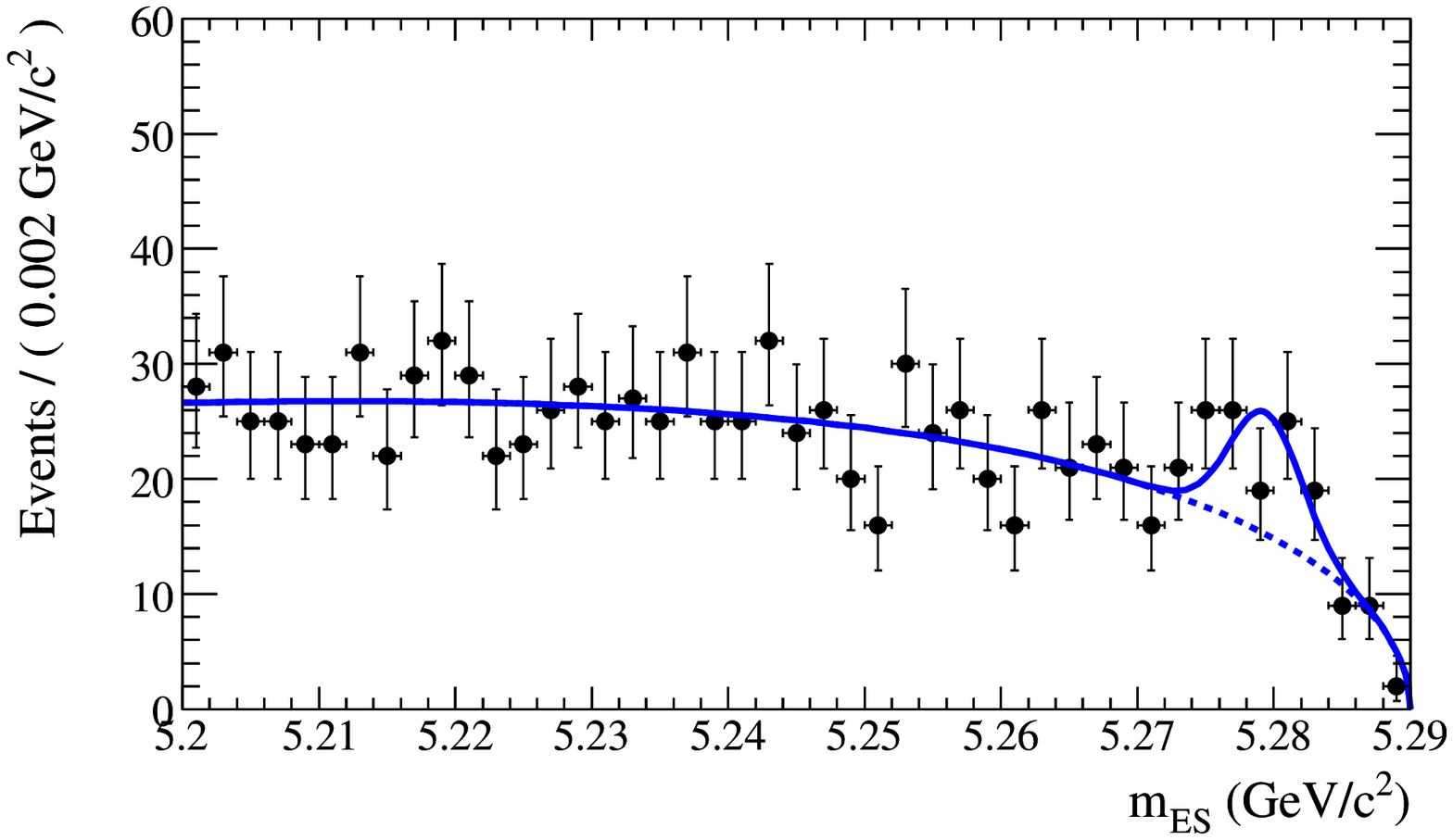}
\put(-170,110){\large ``$D^{+}_s$''$K^- \pi^-$}
\includegraphics[width=8.1cm, height=5.5cm]{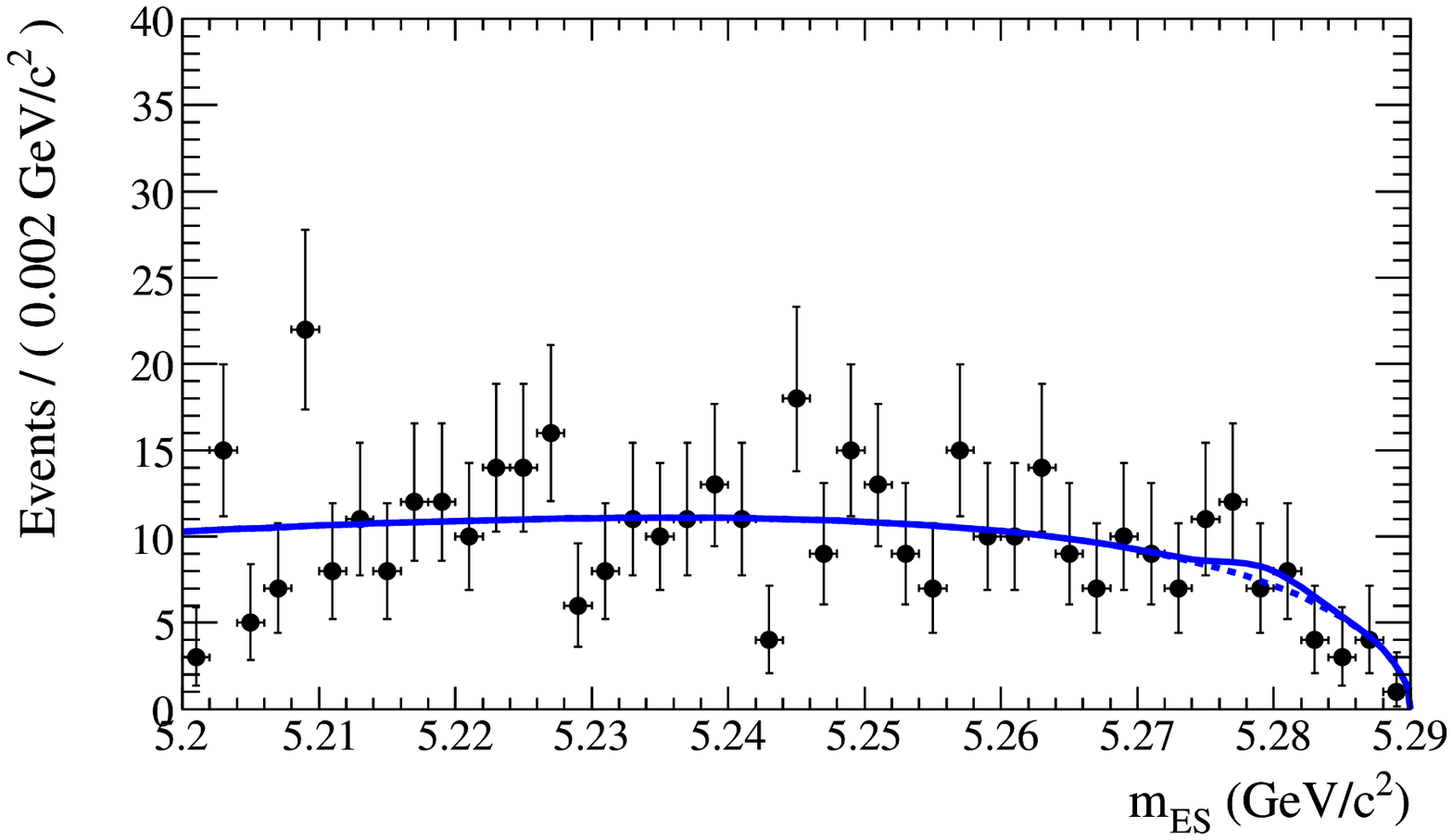}
\put(-170,110){\large ``$D^{*+}_s$''$K^- \pi^-$}
\end{center}
\caption{$m_{ES}$ spectra for the data
``$D^{+}_s$''$K^- \pi^-$ (left) and ``$D^{*+}_s$''$K^- \pi^-$ (right) combinations 
with no mass constraint applied on ``$D^+_s$'' candidates, using $D^+_s$ mass sidebands 
$[\pm 25, \pm 40]$~MeV (1.5 times the signal interval).}
\label{fig:peaking}
\end{figure}

A study using simulated $B^0$ and $B^+$
decays shows that some 
background events with distributions in \mes and in $\Delta E$ peaking
near the signal region
are expected in reconstructed
$\Bm \to D^+_s K^- \pi^-$ candidates due to charmless and charmonium
$\Bm$ decays with the same set of particles in the final state.
For $\Bm \to D^{*+}_s K^- \pi^-$, no background of this
kind is expected, due to the presence of the $\gamma$, which suppresses
charmless and charmonium decay contributions.
The peaking contribution is evaluated using the data
by reconstructing ``$D^{(*)+}_s$''$K^- \pi^-$ combinations, where
``${D^+_s}$'' candidates are selected from $[\pm 40, \pm 25]$~MeV sidebands
around the $D^+_s$ nominal mass.
In this procedure, we use the same selection
requirements, as for the signal, except that ``$D^+_s$'' candidates are
not mass constrained.
The resulting \mes spectra are shown in Figure~\ref{fig:peaking}.
We fit the distributions using an extended unbinned maximum likelihood (ML) fit 
with a sum of a Gaussian (with a width
and central value fixed from the MC simulation) and a threshold
function $f(m_{\rm ES})$ with the floating shape and normalization
(see detailed expression of the likelihood function is Section~\ref{sec:Physics}).
The fit yields $34 \pm 12$ events in the ``signal'' \mes peak for
``$D^+_s$''$K^- \pi^-$ and $3 \pm 7$ for ``$D^{*+}_s$''$K^- \pi^-$.
Since the sideband interval is 1.5 times larger than the $D^+_s$
mass region used for signal selection, this translates
into $23 \pm 8$ peaking background events expected in 
$\Bm \to D^+_s K^- \pi^-$.

We also study cross-feed between the signal modes and other
decays with final states similar to our signal modes, including
$D^{(*)+}_s K^- K^-$. The cut on $\Delta E$ of the $\Bm$
candidates effectively suppresses the cross-feed contributions,
which do not exceed $2\%$ of the reconstructed signal after all the selection
criteria are applied.

\section{SYSTEMATIC STUDIES}
\label{sec:Systematics}

The summary of the systematic uncertainties is presented in Table~\ref{tab:errors}.
The total relative systematic error is estimated to be 
$22\%$ for each $\Bm$ decay mode, with the largest contribution
coming from the $D^+_s$ branching fractions uncertainty ($15\%$). Other significant sources
of systematic errors are found to be due to the difference between the selection
efficiency in MC events and in the data (estimated using the control mode
$\Bm \to D^-_s D^0,~D^0 \to K^- \pi^+$), and also due to the efficiency
dependence on the $D^{(*)+}_s K^-$ invariant mass and its potential effect 
if the resonant contribution is present.

\begin{table}[ht]
{\normalsize
\begin{center}
\caption{\label{tab:errors} Summary of relative systematic errors (in $\%$) 
for $\Bm \ra D_s^{(*)+} K^- \pi^-$ decays.}
\vspace{\baselineskip}
\begin{tabular}{lcc}
Source~~~~~~~~~~~~~~~~~~~~~~~~~~~~~~ & $\Bm \ra D_s^{+} K^- \pi^-$  & 
$\Bm \ra D_s^{*+} K^- \pi^-$ \\[3pt]
\hline\\[-7pt]
$B$ counting  & 1.1 & 1.1 \\[4pt]
MC statistics & 0.8 & 1.4 \\[4pt]
Tracking    & 5 & 5 \\[4pt]
Particle identification efficiency & 4 & 4 \\[4pt]
$K^0_S$ efficiency & 0.5 & 0.5 \\[4pt]
$\gamma$ (from $D^{*+}_s \ra D^+_s \gamma$) efficiency & -- & 2 \\[4pt]
${\cal B}$ of sub-decays & 15 & 15 \\[4pt]
Peaking background contribution & 6 & 3 \\[4pt]
Cross-feed contribution & 1 & 2 \\[4pt]
Selection efficiency, Data/MC & 12 & 12 \\[4pt]
Signal and background shape uncertainty & 3 & 3 \\[4pt]
$M(D^{(*)-}_s K^+)$ efficiency dependence & 7 & 9 \\[4pt]
\hline\\[-7pt]
Total  & 22 & 22 \\[4pt]
\hline
\end{tabular}
\end{center}
}\end{table}

\section{RESULTS}
\label{sec:Physics}

\begin{figure}[ht]
\begin{center}
\includegraphics[width=8.1cm, height=5.5cm]{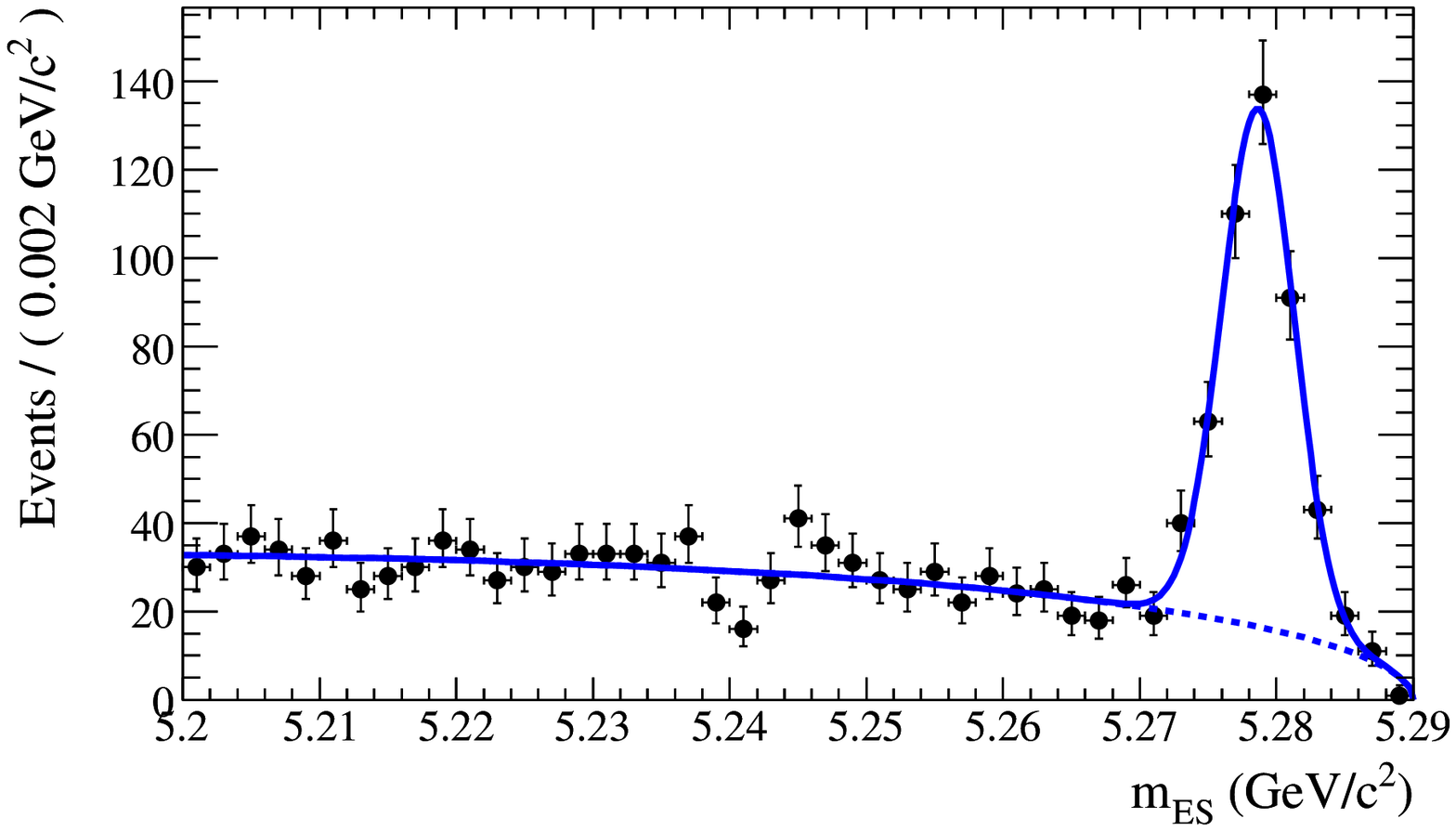}
\put(-180,90){\large $\Bm \to D^+_s K^- \pi^-$}
\includegraphics[width=8.1cm, height=5.5cm]{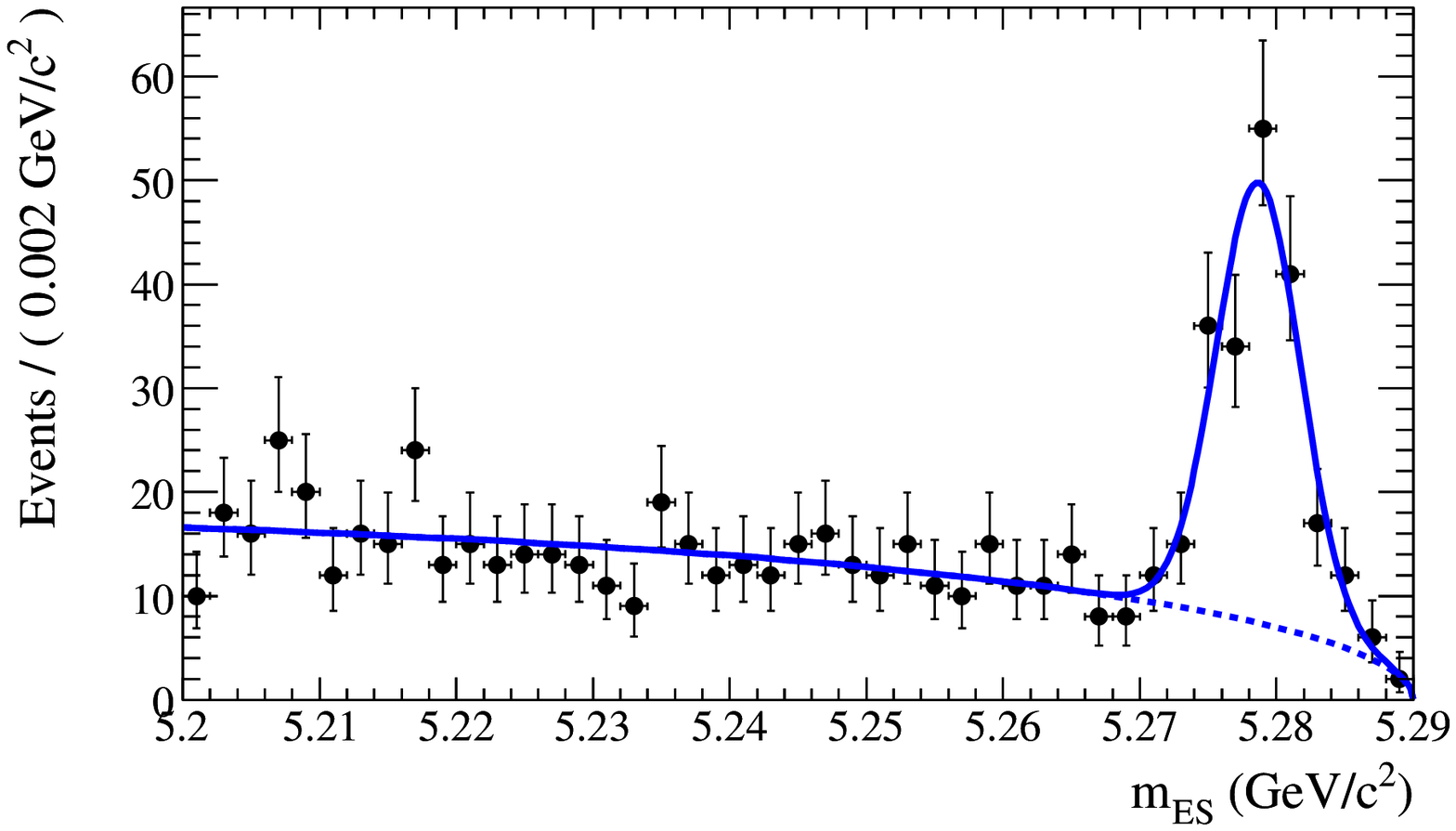}
\put(-180,90){\large $\Bm \to D^{*+}_s K^- \pi^-$}
\end{center}
\caption{\label{fig:mes-all} $m_{ES}$ spectra for the $\Bm \to D^+_s K^- \pi^-$ (left)
and $\Bm \to D^{*+}_s K^- \pi^-$ (right) using the data. Solid curves
show the fit results, as explained in the text. Dashed lines in the signal
regions correspond to the background components of the fit.}
\end{figure}

Figure~\ref{fig:mes-all} shows the \mes
distributions for the reconstructed candidates $\Bm \to D_s^+ K^- \pi^-$ and
$\Bm \to D_s^{*+} K^- \pi^-$.
For each mode, we perform an
extended unbinned ML fit to the \mes distributions using the
candidates from all \Ds decay modes combined. We fit the \mes
distributions with the 
sum of the function $f(m_{\rm ES})$ characterizing
the combinatorial background and a Gaussian function to describe the
signal. 
The mean and width of the Gaussian function,
the threshold shape parameter $\xi$, and the numbers of signal ($n_{sig}$)
and background ($n_{bkg}$) events are free parameters of the fit.
The likelihood function is given by:
$$
{\cal L} = \frac{e^{-(n_{sig}+n_{bkg})}}{N!} \prod_{i=1}^N (n_{sig} P_i^{sig} + n_{bkg} P_i^{bkg}),
\label{eq:likelihood}
$$
where 
$P^{sig}_i$ and $P^{bkg}_i$ are the probability density functions
for the signal and background, $N$ is the total number of events in the fit
and $i$ is the index over all events in the fit.

The fit yields $393 \pm 25$ events in the 
$\Bm \to D^+_s K^- \pi^-$ mode. Taking into acccount the estimated
peaking background contribution, we obtain
$370 \pm 26$ signal events for $\Bm \to D^+_s K^- \pi^-$. The number
of $B^- \to D^{*+}_s K^- \pi^-$ signal events from the fit
is $164 \pm 17$ (no peaking contribution is subtracted in this
mode as it was estimated to be consistent with 0). We also fit signal 
yields in each of the $D^+_s$ modes, fixing
the width and central values of the signal Gaussians to that of the
combined fit, letting the background level and shape float. The ratio of the signal 
yields between submodes is consistent with 
the expectations from MC.

The total signal yield in each $\Bm$ decay mode is calculated as
a sum over $D^+_s$ modes ($i=\phi \pip$, $\Kstarzb K^+$, $\KS K^+$)
and is related to the $\Bm$ branching fraction ${\cal B}$ using the following
expression: 
$$
n_{sig} = {\cal B} \cdot N_{B \bar B} \cdot \sum_i  {\cal B}_i \cdot \epsilon_i, 
$$
where $N_{B \bar B}$ is the number of produced $B\bar B$ pairs,
${\cal B}_i$ is the product of the intermediate
branching ratios and $\epsilon_i$ is the reconstruction efficiency (from Table~\ref{tab:eff}).
As an input to the calculation, we used branching fraction numbers
from~\cite{pdg} and~\cite{dsphipi-babar}. The relative systematic 
uncertainties are
converted into absolute numbers using the measured central
values. The results are:
$$
{\cal B}(\Bm \to D^+_s K^- \pi^-) = (1.88 \pm 0.13 \pm 0.41) \cdot 10^{-4} 
$$
$$
{\cal B}(\Bm \to D^{*+}_s K^- \pi^-) = (1.84 \pm 0.19 \pm 0.40) \cdot 10^{-4}
$$ 

In summary, two decay modes of charged $B$ mesons are observed
for the first time. The significance of the observation is 14.2$\sigma$
for $\Bm \to D^+_s K^- \pi^-$ and 9.6$\sigma$ for $\Bm \to D^{*+}_s K^- \pi^-$.

\section{ACKNOWLEDGMENTS}
\label{sec:Acknowledgments}

We are grateful for the 
extraordinary contributions of our \pep2\ colleagues in
achieving the excellent luminosity and machine conditions
that have made this work possible.
The success of this project also relies critically on the 
expertise and dedication of the computing organizations that 
support \babar.
The collaborating institutions wish to thank 
SLAC for its support and the kind hospitality extended to them. 
This work is supported by the
US Department of Energy
and National Science Foundation, the
Natural Sciences and Engineering Research Council (Canada),
Institute of High Energy Physics (China), the
Commissariat \`a l'Energie Atomique and
Institut National de Physique Nucl\'eaire et de Physique des Particules
(France), the
Bundesministerium f\"ur Bildung und Forschung and
Deutsche Forschungsgemeinschaft
(Germany), the
Istituto Nazionale di Fisica Nucleare (Italy),
the Foundation for Fundamental Research on Matter (The Netherlands),
the Research Council of Norway, the
Ministry of Science and Technology of the Russian Federation, 
Ministerio de Educaci\'on y Ciencia (Spain), and the
Particle Physics and Astronomy Research Council (United Kingdom). 
Individuals have received support from 
the Marie-Curie IEF program (European Union) and
the A. P. Sloan Foundation.


\begin{thebibliography}{99}
\bibitem{charge}
Charge conjugate reactions are implicitly included, throughout this paper.
\relax
\end{thebibliography}
\end{document}